\documentstyle[aps,amssymb,12pt]{revtex}

\begin{document}
\author{S. Bruce$^{\text{(1)}}$\thanks{%
E-mail: sbruce@udec.cl, Fax: 56-41-224520.} , L. Roa$^{\text{(1)}}$ and C.
Saavedra$^{\text{(1)}}$, A.B. Klimov$^{\text{(2)}}$}
\title{Unbroken supersymmetry in the Aharonov-Casher effect}
\address{$^{\text{(1)}}$ Departamento de F\'{i}sica, Universidad de Concepci\'{o}n,\\
Casilla 4009, Concepcion, Chile\\
$^{\text{(2)}}$ Departamento de F\'{i}sica, Universidad de Guadalajara,\\
Corregidora 500, 44420, Guadalajara, Jalisco, M\'{e}xico}
\date{April, 1999}
\maketitle

\begin{abstract}
We consider the problem of the bound states of a spin $1/2$ chargless
particle in a given Aharonov-Casher configuration. To this end we recast the
description of the system in a supersymmetric form. Then the basic physical
requirements for unbroken supersymmetry are established. We comment on the
possibility of neutron confinement in this system.

PACS number(s): 03.65Ge, 03.65.Bz, 12.60.J, 11.30.P.
\end{abstract}

Aharonov and Casher \cite{AC} introduced a `dual' to the well established
Aharonov-Bohm effect \cite{AB}. The essence of the Aharonov-Bohm effect is
the presence of the vector potential in the Lagrangian formulation used in
quantum mechanics. A charged particle moving through a region close to a
magnetic field experiences no Lorentz force but is modified by a non-zero
vector potential in the equation of motion \cite{AB,PE,H,XI}. Based on the
`symmetry' of Maxwell's equations Aharonov and Casher considered the
interaction between a particle's magnetic dipole moment and an electric
field. A fully relativistic theory of the Aharonov-Casher effect has been
given by Hagen \cite{H} and He and McKellar \cite{XI} for spin $1/2$
particles .

The AC phase shift was observed using neutron interferometry \cite{K,GO,KA}.
In the same year that the Aharonov-Casher effect was announced, M. V. Berry
introduced the concept of the geometric or topological phase in quantum
mechanics \cite{WI}. In cases where the adiabatic theorem can be invoked, a
non-integrable (i.e. non-dynamic) phase is accumulated in the cyclic
evolution of a Hamiltonian which is not simply connected. An important
example of this geometric phase was the Aharonov-Bohm effect. Although
classical examples have been found the topological nature of the AB and AC
phase is an important argument for their quantum nature.

A significant investigation of the ``duality'' of AB and AC effects has been
carried out to show the equivalence of these effects by transformation of
one equation of motion into the other \cite{H,SA,AZ}. Other investigations
of the Aharonov-Casher effect have included extensions into massive photon
electrodynamics \cite{FU}, non-locality \cite{RE}, abstract geometry \cite
{MI}, gravity \cite{RE2}, and AB/AC interference \cite{XG,LE}.

Here we are concerned with another application of the AC effect. It deals
with the conditions for finding the bound states of a system with {\it %
unbroken} supersymmetry. To this end we have to assume {\it connectness} in
the configuration space in order to be able to define a normalizable ground
state. The problem turns out to have exact supersymmetry only under the
fulfillment of a condition for the magnitude of the charge distribution
which generates the electric field. We also discuss the possibility of {\it %
breaking} supersymmetry by examining the requirements for the existence of
lower energy bound states.

To start with, let us consider an infinite cylinder with uniform charge per
unit volume $\rho $ centered along the $z$ axis, so that there exists an
electric field 
\begin{equation}
{\bf E}_{<}{\bf (r)}=\rho {\bf r}/2,{\bf \qquad }0{\bf \leq }r\leq
r_{0};\qquad {\bf E}_{>}{\bf (r)}=\rho r_{0}^{2}{\bf r}/2r^{2}{\bf ,\qquad }%
r_{0}\leq r<\infty ,  \label{E}
\end{equation}
where $r_{0}$ is the radius of the cylinder and for simplicity we have
chosen $\widehat{{\bf r}}{\bf \cdot \widehat{z}=}0.$ Here $\widehat{{\bf r}}$
and $\widehat{{\bf z}}$ are unit vectors in the ${\bf r}$ and ${\bf z}$
directions respectively. The uncharged particles (v.gr. neutrons) are
completely polarized along the positive ${\bf z}$ direction. They move on a
plane in the external field ${\bf E.}$ In this circumstance there is
apparently no force on the neutrons but there exists a kind of Aharonov-Bohm
effect \cite{AC,H,K}. Nevertheless, if the singularity in the $z$ axis is
removed, as is implied in (\ref{E}), the neutrons are allowed to penetrate
the charged line. Therefore a new question is to be considered: It regards
the problem of the possible bound states of the neutron in this new AC
configuration.

To be specific, let us consider a spin $1/2$ chargless particle with an
anomalous magnetic moment $\kappa _{n}.$ The Dirac equation can be written 
\cite{AC,H} in a covariant form ($\hbar =c=1)$ as 
\begin{equation}
\left( \gamma _{\mu }p^{\mu }-\frac{e\kappa _{n}}{2M_{n}}F^{\mu \nu }\sigma
_{\mu \nu }-M_{n}\right) \Psi ({\bf r,}t)=0,  \label{Pa}
\end{equation}
where $F^{\mu \nu }=\partial ^{\mu }A^{\nu }-\partial ^{\nu }A^{\mu }$ is
the electromagnetic field tensor.

The Aharonov-Casher effective wave equation is obtained by making $A_{0}\neq
0,${\bf \ }${\bf B}={\bf 0,}$ with ${\bf \nabla \cdot E=}\rho $. For
stationary states of energy $E$ we write 
\begin{equation}
\Psi _{E}({\bf r,}t)=%
{\phi ({\bf r}) \choose \chi ({\bf r})}%
e^{-iEt}.  \label{DWF}
\end{equation}
Thus from (\ref{Pa}) and (\ref{DWF}) we get 
\begin{eqnarray}
\frac{1}{2M_{n}}{\bf \sigma }\cdot \left( {\bf p+}i\eta {\bf E(r)}\right) 
{\bf \sigma }\cdot \left( {\bf p-}i\eta {\bf E(r)}\right) \phi ({\bf r}) &=&%
\frac{\varepsilon }{2M_{n}}\phi ({\bf r}),  \label{ul} \\
\frac{1}{2M_{n}}{\bf \sigma }\cdot \left( {\bf p-}i\eta {\bf E(r)}\right) 
{\bf \sigma }\cdot \left( {\bf p+}i\eta {\bf E(r)}\right) \chi ({\bf r}) &=&%
\frac{\varepsilon }{2M_{n}}\chi ({\bf r}),  \nonumber
\end{eqnarray}
where ${\bf \sigma =(}\sigma _{1},\sigma _{2})$, $\eta =e\kappa _{n}/2M_{n}$
and $\varepsilon \equiv E^{2}-M_{n}^{2}.$ This set of {\it uncoupled}
differential equations can be rewritten in the supersymmetric form 
\begin{equation}
{\rm H}_{SS}{\cal =}\left\{ Q,Q^{\dagger }\right\} ,\qquad \left[ {\rm H}%
_{SS},Q\right] =\left[ {\rm H}_{SS},Q^{\dagger }\right] =0,  \label{HAA}
\end{equation}
with 
\begin{equation}
{\rm H}_{SS}\Psi _{E}({\bf r,}t)=\frac{\varepsilon }{2M_{n}}\Psi _{E}({\bf r,%
}t).  \label{Su}
\end{equation}
Here 
\begin{equation}
Q\equiv \frac{1}{\sqrt{2M_{n}}}\tau ^{-}\otimes {\bf \sigma }\cdot \left( 
{\bf p-}i\eta {\bf E(r)}\right)  \label{Sq}
\end{equation}
is the supersymmetric charge and $\tau ^{-}=(1/2)\left( \tau _{1}-i\tau
_{2}\right) ,$ where the $\tau _{1},$ $\tau _{2}$ are Pauli matrices. Thus $%
{\rm H}_{SS}$ is invariant under $N=1$ supersymmetry. From (\ref{ul}) we
find \cite{BR} that 
\begin{equation}
\left\{ {\bf p}^{2}+\eta \tau _{3}\otimes \left( \nabla {\bf \cdot E(r)}%
+2\sigma _{3}\left( {\bf E(r)\times p}\right) _{3}\right) +\eta ^{2}{\bf E}%
^{2}({\bf r})\right\} \Psi ({\bf r})=\varepsilon \Psi ({\bf r}).  \label{Sy}
\end{equation}

It is not surprising to find here a supersymmetric system since the
Hamiltonian in (\ref{HAA}) describes the interaction between a spin 1/2
particle with an electromagnetic field (spin 1). Note that the AC effect has
also been discussed in the framework of $N=2$ nonrelativistic supersymmety 
\cite{RO}.

Supersymmetry is unbroken if 
\begin{equation}
Q\phi ^{(0)}({\bf r})=0,\qquad Q^{\dagger }\phi ^{(0)}({\bf r})=0,
\label{Uc}
\end{equation}
where $\phi ^{(0)}$ is the ground state of the system. In other words, the
generators of $N=1$ supersymmetry annihilate the vacuum state in order to
have an exact symmetry. Furthermore, in a system with {\it axial} symmetry
we have also the constraint 
\begin{equation}
\left( {\bf E(r)\times p}\right) _{3}\phi ^{(0)}({\bf r})=\frac{\mid {\bf %
E(r)\mid }}{r}L_{3}\phi ^{(0)}({\bf r})=0\qquad \text{ }(\text{a }s\text{%
-state}),  \label{Con}
\end{equation}
with $L_{3}{\bf =}\left( {\bf r\times p}\right) _{3}$ the $z$ component of
the orbital angular momentum operator. Here then we are concerned with
states for which $E^{2}=M_{n}^{2}$, i.e., $\varepsilon =0$.

The second equation (\ref{Uc}) is satisfied identically since in the
nonrelativistic limit the lower components $\Psi _{E=M_{n}}$ vanish. The
first one together with (\ref{Con}) yield 
\begin{equation}
{\bf \sigma }\cdot \left( {\bf p-}i\eta {\bf E(r)}\right) \phi ^{(0)}(r)=0.
\label{Co}
\end{equation}
Without lack of generality we can set 
\begin{equation}
\phi ^{(0)}(r)\equiv 
{\phi (r) \choose 0}%
,\qquad \chi ^{(0)}(r)\equiv 
{0 \choose 0}%
.
\end{equation}
Then from (\ref{Co}) we find the first order differential equations 
\begin{equation}
\left( \frac{d}{dr}{\bf -}\beta r\right) \phi _{<}(r)=0,\qquad 0\leq r\leq
r_{0};\qquad \left( \frac{d}{dr}{\bf -}\frac{\beta r_{0}^{2}}{r}\right) \phi
_{>}(r)=0,\qquad r_{0}\leq r<\infty ,  \nonumber
\end{equation}
where $\beta \equiv \rho \eta /2.$ Thus 
\begin{equation}
\phi _{<}(r)=Ae^{-\beta r^{2}/2},\qquad 0\leq r\leq r_{0};\qquad \phi
_{>}(r)=Br^{-\beta r_{0}^{2}},\qquad r_{0}\leq r<\infty ,  \label{sst}
\end{equation}
with $A,B$ complex constants.

Next we demand continuity of the wavefunction and its derivative at $r=r_{0}.
$ Both conditions give the same information$:$%
\begin{equation}
Ae^{-\beta r_{0}^{2}/2}=Br_{_{0}}^{-\beta r_{0}^{2}}.  \label{AB}
\end{equation}
Furthermore, if $\Psi _{E=M_{n}}$ belongs to the Hilbert space, $\phi $ must
be normalizable on the plane $[0,2\pi ]\times \lbrack 0,\infty \lbrack $: 
\begin{equation}
2\pi \int_{0}^{\infty }\mid \phi (r)\mid ^{2}rdr=2\pi \left\{ \mid A\mid
^{2}\int_{0}^{r_{0}}drre^{-\beta r^{2}}+\mid B\mid ^{2}\int_{r_{0}}^{\infty
}drr^{-2\beta r_{0}^{2}+1}\right\} =1.  \label{N}
\end{equation}
By using (\ref{AB}) we get 
\begin{equation}
\mid A\mid ^{2}=\frac{\beta \left( \beta r_{0}^{2}-1\right) e^{\frac{1}{2}%
\beta r_{0}^{2}}}{2\pi \left[ \left( \beta r_{0}^{2}-1\right) \sinh \left( 
\frac{1}{2}\beta r_{0}^{2}\right) +\frac{1}{2}\beta r_{0}^{2}e^{-\frac{1}{2}%
\beta r_{0}^{2}}\right] }\ .  \label{No}
\end{equation}
Notice that in (\ref{N}) we must require that 
\begin{equation}
\beta r_{0}^{2}>1.  \label{Be}
\end{equation}
This inequality constitutes a constraint on the possible values of $\rho $
and $r_{0}$ (or equivalently for $\lambda \equiv \rho \pi r_{0}^{2}$) if we
want to preserve unbroken supersymmetry. Inserting $c^{2}$ in (\ref{Be}), we
can estimate the minimum value of $\lambda $ to be able to obtain a
normalizable ground state: $\left| \lambda \right| _{\min }\backsimeq 4\pi
M_{n}c^{2}/\left| e\kappa _{n}\right| \backsimeq 20.\,62$ $\times 10^{-3}$
[C/cm]. As $\lambda $ depends linearly on $r_{0}^{2}$, one can in principle
set up a configuration with the required $\lambda .$

Figure $1$ shows the neutron density of probability $\mid \phi \mid ^{2}$ as
a function of the dimensionless parameter $r/r_{0}$ for different values of $%
\beta >1,\ $in natural units. Notice that when $\beta $ approaches $1$, $%
\mid \phi \mid ^{2}$ becomes flatter, i.e., there exists a larger
probability that the neutron be outside the charged distribution than within
it. This is also easily represented by the ratio of probabilities 
\begin{equation}
R_{\beta }\equiv \frac{W[r_{0},\infty \lbrack }{W[0,r_{0}]}=\frac{\beta
r_{0}^{2}}{2\left( \beta r_{0}^{2}-1\right) \sinh \left( \frac{1}{2}\beta
r_{0}^{2}\right) e^{\frac{1}{2}\beta r_{0}^{2}}}\ ,
\end{equation}
where 
\begin{equation}
W[r_{1},r_{2}]=2\pi \int_{r_{1}}^{r_{2}}\mid \phi (r)\mid ^{2}rdr,
\end{equation}
for values of $\beta >1/r_{0}^{2}$ (fig. $2$).

The general eigenvalue problem for (\ref{Sy}) reduces to the system of
differential equations 
\begin{eqnarray}
\left( -{\bf \nabla }^{2}{\bf -}\eta \left( {\bf \nabla \cdot E(r)}\pm
2\left( {\bf E(r)\times p}\right) _{3}\right) +\eta ^{2}{\bf E}^{2}({\bf r}%
)\right) \phi ^{(%
{1 \atop 2}%
)}({\bf r}) &=&\varepsilon \phi ^{(%
{1 \atop 2}%
)}({\bf r}),  \nonumber \\
&&  \label{Sy11} \\
\left( -{\bf \nabla }^{2}{\bf +}\eta \left( {\bf \nabla \cdot E(r)}\pm
2\left( {\bf E(r)\times p}\right) _{3}\right) +\eta ^{2}{\bf E}^{2}({\bf r}%
)\right) \chi ^{(%
{1 \atop 2}%
)}({\bf r}) &=&\varepsilon \chi ^{(%
{1 \atop 2}%
)}({\bf r}).  \nonumber
\end{eqnarray}
We solve (\ref{Sy11}) by separation of variables: 
\begin{equation}
\phi ({\bf r})=\phi (r)\exp (im\varphi ),\qquad \chi ({\bf r})=\chi (r)\exp
(im\varphi ).  \label{SE}
\end{equation}
Thus from (\ref{Sy11}) and (\ref{SE}) we get 
\begin{eqnarray}
\left( -\frac{d^{2}}{dr^{2}}-\frac{1}{r}\frac{d}{dr}+\frac{m^{2}}{r^{2}}%
-2\beta \left( 1\pm m\right) +\beta ^{2}r^{2}\right) \phi _{<}^{^{(%
{1 \atop 2}%
)}}(r) &=&\varepsilon \phi _{<}^{^{(%
{1 \atop 2}%
)}}(r),  \nonumber \\
\left( -\frac{d^{2}}{dr^{2}}-\frac{1}{r}\frac{d}{dr}+\frac{m^{2}}{r^{2}}%
+2\beta \left( 1\pm m\right) +\beta ^{2}r^{2}\right) \chi _{<}^{^{(%
{1 \atop 2}%
)}}(r) &=&\varepsilon \chi _{<}^{^{(%
{1 \atop 2}%
)}}(r),  \nonumber
\end{eqnarray}
for $r\leq r_{0}$, and 
\begin{eqnarray}
\left( -\frac{d^{2}}{dr^{2}}-\frac{1}{r}\frac{d}{dr}+\frac{m\left( m\mp
2\beta r_{0}^{2}\right) }{r^{2}}+\left( \frac{\beta r_{0}^{2}}{r}\right)
^{2}\right) \phi _{>}^{^{(%
{1 \atop 2}%
)}}(r) &=&\varepsilon \phi _{>}^{^{(%
{1 \atop 2}%
)}}(r),  \nonumber \\
\left( -\frac{d^{2}}{dr^{2}}-\frac{1}{r}\frac{d}{dr}+\frac{m\left( m\pm
2\beta r_{0}^{2}\right) }{r^{2}}+\left( \frac{\beta r_{0}^{2}}{r}\right)
^{2}\right) \chi _{>}^{^{(%
{1 \atop 2}%
)}}(r) &=&\varepsilon \chi _{>}^{^{(%
{1 \atop 2}%
)}}(r),  \nonumber
\end{eqnarray}
for $r\geq r_{0}$.

The radial solutions must be normalizable in the range $0\leq r<\infty $,
and we also demand continuity at $r_{0}$ on the corresponding solutions$.$
For the upper component $\phi _{<}^{(1)}(r)$ we get 
\begin{equation}
\left( \frac{d^{2}}{dr^{2}}+\frac{1}{r}\frac{d}{dr}-\frac{m^{2}}{r^{2}}%
-\beta ^{2}r^{2}+{\sl \epsilon }_{m}\right) \phi _{<}^{(1)}(r)=0,
\end{equation}
$\allowbreak $where $\epsilon _{m}\equiv \varepsilon +2\beta \left(
1+m\right) $. Thus 
\begin{equation}
\phi _{<}^{(1)}(r)=C\,_{1}F_{1}\left( \frac{m+1-\epsilon _{m}/2\beta }{2}%
;m+1;2\beta r^{2}\right) r^{m}e^{-\beta r^{2}/2},  \label{fi1}
\end{equation}
where $C$ is a complex constant and $_{1}F_{1}$ is the {\it confluent
hypergeometric function}.

For $r\geq r_{0}$ we have 
\begin{equation}
\left( \frac{d^{2}}{dr^{2}}+\frac{1}{r}\frac{d}{dr}-\frac{l^{2}}{r^{2}}%
+\varepsilon \right) \phi _{>}^{(1)}(r)=0  \label{mr}
\end{equation}
where $l\equiv m-\beta r_{0}^{2}$. Eq.(\ref{mr}) has two kinds of solutions:
(a) non-normalizable scattering-like states for $\varepsilon >0$ ( $%
E^{2}>M_{n}^{2}$ ): 
\begin{equation}
\phi _{>}^{(1)}(r)=C_{1}J_{l}\left( \sqrt{\varepsilon }r\right)
+C_{2}Y_{l}\left( \sqrt{\varepsilon }r\right) ,\qquad \left| m\right| >1,
\end{equation}
where $J_{l}(z)$ and $Y_{l}(z)$ are the Bessel functions of first and second
kind correspondingly; and (b) bound states for $\varepsilon <0$ ( $%
E^{2}<M_{n}^{2}$ ):

\begin{equation}
\phi _{>}^{(1)}(r)=DK_{l}\left( \sqrt{-\varepsilon }r\right) ,\qquad \left|
m\right| \geq 0,  \label{fi2}
\end{equation}
where $K_{l}(z)$ is the (normalizable) modified Bessel function of second
kind. By matching (\ref{fi1}) and (\ref{fi2}) at $r=r_{0},$ we find the
quantization condition for the remaining bound state energy levels. This
problem is now under research. We note that the supersymmetric state (\ref
{sst}) is obtained as a limit case ($\varepsilon \longrightarrow 0$ for $m=0$%
) of bound state solutions of the form (\ref{fi2}). In fact, in this limit $%
\phi _{<}^{(1)}(r)\varpropto e^{-\beta r^{2}/2}$. The existence of further
bound states would break exact supersymmetry (stated by (\ref{HAA}) and (\ref
{Uc})), since $\varepsilon _{{\rm \min }}<0$.

From the above we can draw at least two main conclusions. First, the
magnitude of the electric charge distribution has to be sufficiently large $%
\left( \lambda \gtrsim 4\pi M_{n}c^{2}/\left| e\kappa _{n}\right| \right) $
in order to generate a bound (ground) state. Second, we are not asserting
that the neutron {\it directly }``feels'' a force due to the electric field
generated by the charge density. Rather, from the second term of the left
hand side of (\ref{Sy}), we state that the neutron tends to move toward
regions where the {\it gradient} of the electric field increases. The third
term in the same equation corresponds to the appearance of an induced
electric dipole moment on the particle \cite{AC}.

It is interesting to note that the fulfillment of condition (\ref{Be}) would
allow neutron trapping by an electrostatic field as a result of a purely
quantum mechanical effect.

{\bf Acknowledgments}

This work was supported by Direcci\'{o}n de Investigaci\'{o}n, Universidad
de Concepci\'{o}n, through grants P.I. 96.11.19-1.0 and Fondecyt \#1970995.

One of us (SB) is grateful to the School of Physics, University of
Melbourne, Australia, for its warm hospitality. We are very thankful to
Professors A. G. Klein and G. I. Opat for their valuable criticisms and
helpful suggestions on the experimental and theoretical aspects of this
paper.\newpage

{\LARGE Figure Captions\bigskip }

FIG. 1. The neutron ground state probability density $\mid \phi \left(
r\right) \mid ^{2}$ as a function of the dimensionless parameter $r/r_{0}$
for different values of $\beta >1$. The units used are $\hbar =c=1$.\bigskip

FIG. 2. The ratio of probabilities $R_{\beta }\equiv W[r_{0},\infty \lbrack
/W[0,r_{0}]$ for values of the parameter $\beta >1$ with $r_{0}=1.$ Notice
that the integration area under each of the three curves is different.
However, $W[0,\infty \lbrack =1$ for each one of them, since the integration
is performed on the plane $[0,2\pi ]\times \lbrack 0,\infty \lbrack $ where
the measure is $2\pi rdr.$

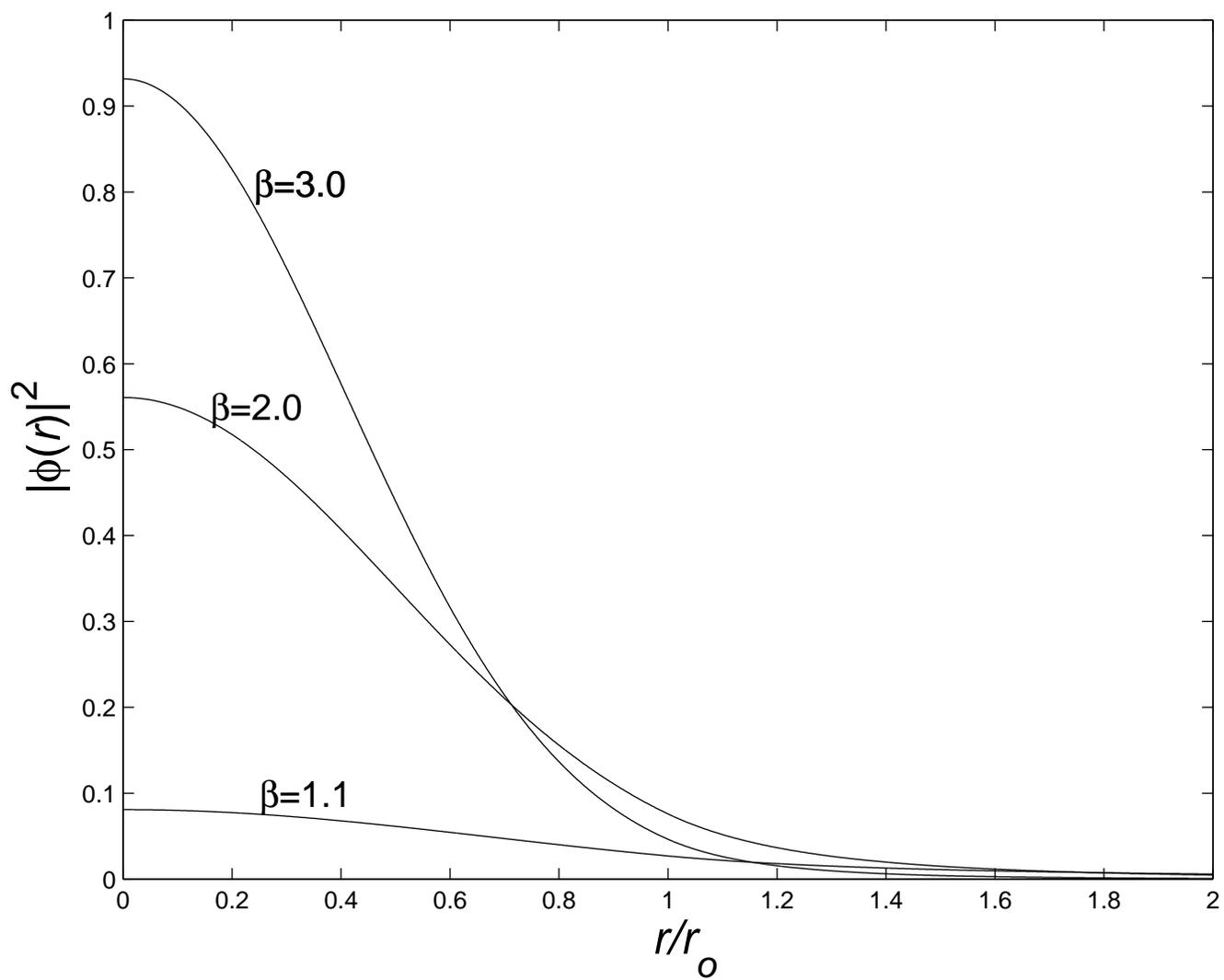
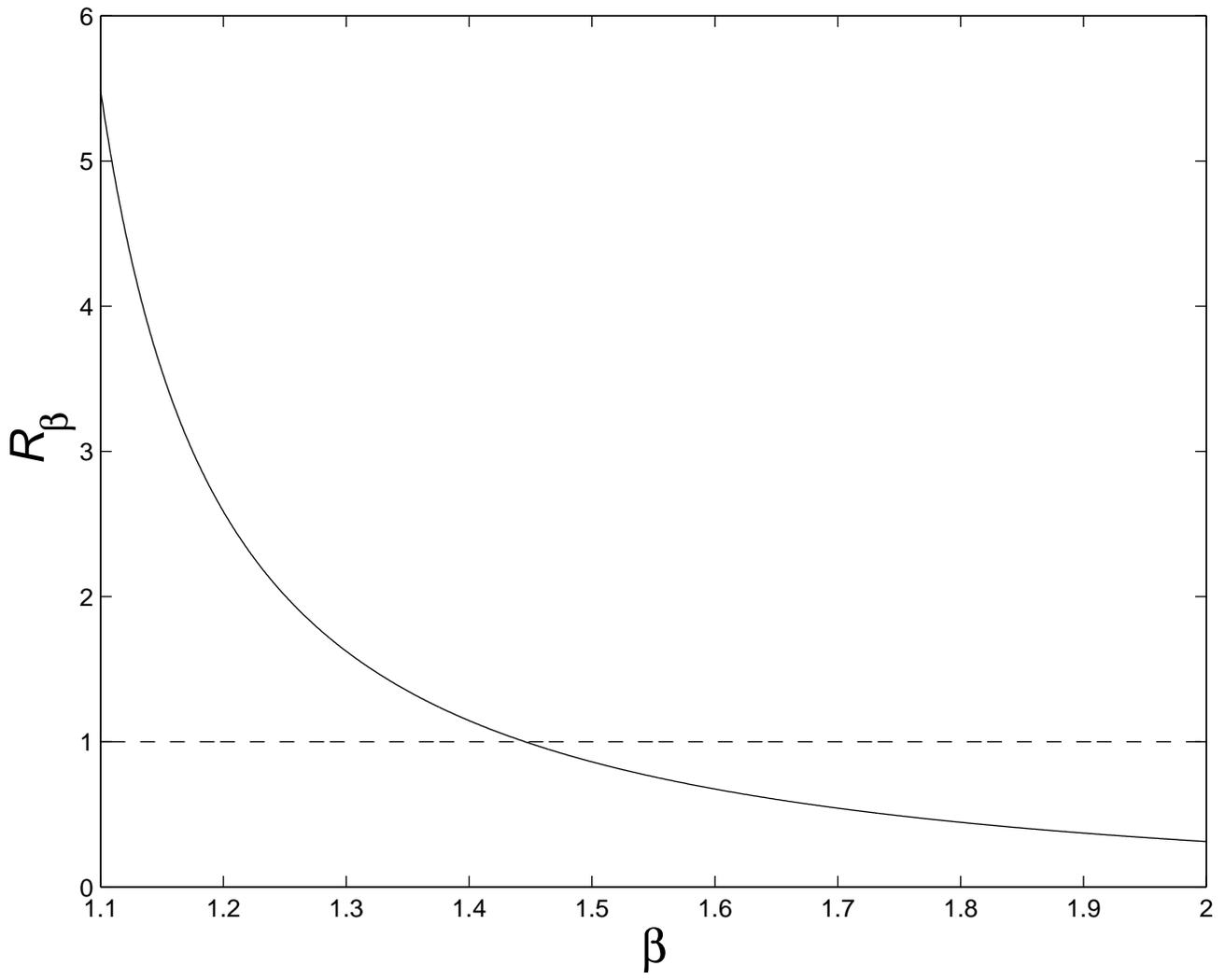
\end{document}